\pdfoutput=1
\documentclass[twocolumn,english,superscriptaddress,footinbib,bibnotes]{revtex4-1}

\usepackage{graphicx}
\usepackage[usenames,dvipsnames]{color}
\usepackage{dcolumn}
\usepackage{bbm}
\usepackage{bm}
\usepackage{amsmath}
\usepackage{amssymb}
\usepackage{times}
\usepackage{placeins}
\usepackage{hyperref}

\hypersetup{colorlinks=false}

\newcommand{\vareps}{\varepsilon}
\graphicspath{{Figures/}}

\begin{document}

\title{Quantum phase recognition via unsupervised machine learning}
\date{\today}

\author{Peter Broecker}
\affiliation{Institute for Theoretical Physics, University of Cologne, 50937 Cologne, Germany}
\author{Fakher F. Assaad}
\affiliation{Institut f\"ur Theoretische Physik und Astrophysik, Universit\"at W\"urzburg, 97074 W\"urzburg, Germany}
\author{Simon Trebst}
\affiliation{Institute for Theoretical Physics, University of Cologne, 50937 Cologne, Germany}


\begin{abstract}
The application of state-of-the-art machine learning techniques to statistical physic problems has 
seen a surge of interest for
their ability to discriminate phases of matter by extracting essential features in the many-body wavefunction
or the ensemble of correlators sampled in Monte Carlo simulations.
Here we introduce a generalization of supervised machine learning approaches 
that allows to accurately map out phase diagrams of interacting many-body systems without any prior knowledge,
e.g.~of their general topology or the number of distinct phases.
To substantiate the versatility of this approach, which combines convolutional neural networks with quantum Monte Carlo sampling,
we map out the phase diagrams of interacting boson and fermion models both at zero and finite temperatures and show that first-order, 
second-order, and Kosterlitz-Thouless phase transitions can all be identified. 
We explicitly demonstrate that our approach is capable of identifying the phase transition to non-trivial many-body phases
such as superfluids or topologically ordered phases without supervision.
\end{abstract}

\maketitle

\noindent
In statistical physics, a continuous stream of computational and conceptual advances has been directed towards attacking the
 quantum many-body problem  -- the identification of the ground state of a macroscopic number of interacting bosons, spins or fermions.
Pivotal steps forward have included the development of numerical many-body techniques such as quantum Monte Carlo simulations \cite{Gubernatis2016}
and the density matrix  renormalization group \cite{White1992,Schollwoeck2005} along with conceptual advances such as the formulation
of an entanglement perspective \cite{Amico2008,Eisert2010} on the quantum many-body problem arising from the interplay of quantum information theory
and quantum statistical physics.
Currently, machine learning (ML) approaches are entering this field as new players.
Their core functions, dimensional reduction and feature extraction, are a perfect match to the goal of identifying essential characteristics
of a quantum many-body system, which are often hidden in the exponential complexity of its many-body wavefunction or the abundance of potentially
revealing correlation functions.
Initial steps in this direction have demonstrated that machine learning of wave functions is indeed possible \cite{Carleo2017,Torlai2017},
which can lead to a variational representation of quantum states based on artificial neural networks that, for some cases, outperforms
entanglement-based variational representations \cite{Carleo2017}.
This ability of machine learning algorithms to learn complex distributions has also been utilized to improve Monte Carlo sampling techniques
\cite{Huang2017,Liu2017} and might point to novel ways to bypass the sign problem of the many-fermion problem \cite{Broecker2016}.
In parallel, it has been demonstrated that convolutional neural networks can be trained to learn sufficiently many features from the correlation
functions of a classical many-body system such that distinct phases of matter can be discriminated and the parametric location of the phase
transition between them identified \cite{Carrasquilla2017}.
This supervised learning approach has been generalized to quantum many-body systems \cite{Broecker2016,Khatami2016}, for which the application of additional preprocessing filters even allows for the identification of topological order \cite{Zhang2017,Zhang2017b}.

\begin{figure}[b]
	\includegraphics[width=0.9\columnwidth]{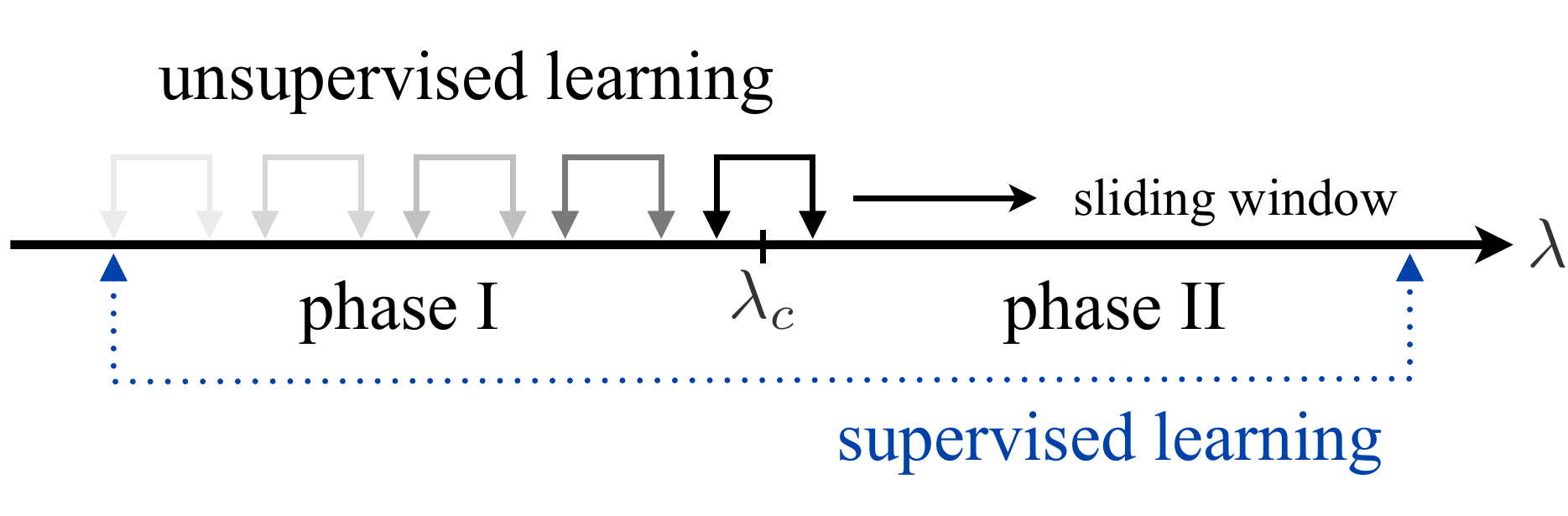}
	\caption{Schematic illustration of the {\bf unsupervised machine learning approach}.
		 	For a small parameter window, which is slided across parameter space, a discrimination of phases at its endpoints A and B
			is attempted via a supervised learning approach. A positive discrimination via the underlying convolutional neural network
			is expected only if the parameter window indeed encompasses a phase transition, while it should fail when points A and B reside
			in the same phase.}
	\label{Fig:Concept}
\end{figure}
In this manuscript, we introduce an unsupervised machine learning approach to the quantum many-body problem that is capable
of parametrically mapping out phase diagrams. The algorithm, which generalizes previous supervised learning schemes to distinguish
phases of matter, works without any prior knowledge, e.g.~regarding the overall topology or number of distinct phases present in a
phase diagram.
The essential ingredient of our approach are convolutional neural networks (CNN) \cite{Nielsen2016} that combine a preprocessing step using convolutional filters with a conventional neural network (typically involving multiple layers itself).
In previous work \cite{Carrasquilla2017,Broecker2016,Khatami2016,Zhang2017,Zhang2017b} such CNNs have been used in a {\em supervised}
learning setting where a (quantum) many-body Hamiltonian is considered that, as a function of some parameter $\lambda$, exhibits a phase transition between two phases -- such as the thermal phase transition in the classical Ising model \cite{Carrasquilla2017} or
the zero-temperature quantum phase transition as a function of some coupling parameter \cite{Broecker2016}.
In such a setting where one has prior knowledge about the existence of two distinct phases in some parameter range, one can train the CNN with {\em labeled} configurations or Green's functions acquired deep inside the two phases (e.g.~by Monte Carlo sampling). After successful training the CNN to distinguish these two phases (which typically requires a few thousand training instances), one can then feed unlabeled instances, sampled for arbitrary intermediate parameter values of $\lambda$, to the CNN in order to locate the phase transition between the two phases,
see also the schematic illustration of Fig.~\ref{Fig:Concept}. This approach has been demonstrated to produce relatively good quantitative estimates for the location of the phase transition \cite{Carrasquilla2017,Broecker2016,Khatami2016,Zhang2017,Zhang2017b}
and might even be finessed to be amenable to a finite-size scaling analysis for second-order phase transitions \cite{Carrasquilla2017}.
This demonstrates that CNNs not only master discrete classification problems as typically encountered in image classification
problems \cite{Nielsen2016}, but are similarly well suited to work on {\em continuous} classification problems
with a CNN exposed to gradually changing instances that nevertheless exhibit one singular (phase) transition that is correctly identified.


\noindent
{\em Unsupervised learning.--}
An essential element in the supervised learning approach outlined above is that the CNN is indeed capable of discriminating two phases of matter
after training it on labeled instances representing the two phases.
It is precisely this realization of a positive discrimination that allows one to construct an {\em unsupervised} machine learning approach to the phase recognition problem encountered in (quantum) many-body problems.
The key idea is to swipe a small window through the parameter space of the to-be-determined phase diagram
and to test whether it is possible to positively discriminate two distinct phases for the two boundary parameters.
The latter can be accomplished by employing the original supervised approach, i.e. by making an attempt to train a CNN on labeled instances for the two boundary parameters and to subsequently determine whether the training indeed led to a positive discrimination of the two sets of instances.
For a parameter window that encloses a phase transition, the expectation is that a positive discrimination is indeed found, i.e. the CNN produces prediction values of $p=0$ and $p=1$ for the two sets of instances, respectively.
For a parameter window that for its full extent resides within one given phase, however, the attempted training should not allow for a positive discrimination of the instances and should result in a prediction value of $p=0.5$ for all instances (indicating a maximal confusion).
This approach is indeed unsupervised in the sense that one can introduce a metric that quantifies how well the instances used in a given training procedure can be positively discriminated. Specifically, we consider the ``label distance" $d(\lambda_1, \lambda_2)$ as the integral over the prediction values $p$ in $\lambda$-space between two points $\lambda_1$ and $\lambda_2$
\begin{equation}
    d(\lambda_1, \lambda_2) = \Theta\left(\; \int\limits_{\lambda_1}^{\lambda_2} \text{d}\lambda (p(\lambda) - 0.5) - \vareps\right)\,,
\end{equation}
which indicates how close the assignment of labels for instances at these two points should be.
If they belong to the same phase, the prediction will in theory always be $0.5$ and the integral will evaluate to $0$, while it will be $1$ if the instances for the two values of $\lambda$ are distinguishable (and separated by a phase transition).
In practice, deviations from the ideal values can be accounted for by introducing a threshold difference $\vareps$ in the above definition.

\noindent
{\em Convolutional neural networks.--}
We implemented \cite{TensorFlow} our method using a CNN comprised of one convolutional layer with 32 filters of size 3x3 
which, after applying a pooling operation, are fed into a fully connected layer of 512 neurons.
A dropout regularization to prevent overfitting is applied before two softmax neurons are used to signal the result.
Except for the output neurons, all neurons are activated by ReLU functions.
As a loss function, we used the cross entropy and an additional $L_2$-regularization on the the weights of the fully connected layer 
with prefactor $\beta = 0.01$.
The optimization of the weights was performed using the ADAM optimizer \cite{Kingma2014}. 
For training the network, we used some 4000 instances and equally many for the prediction in the supervised approach.


\begin{figure}[t]
	\includegraphics[width=0.9\columnwidth]{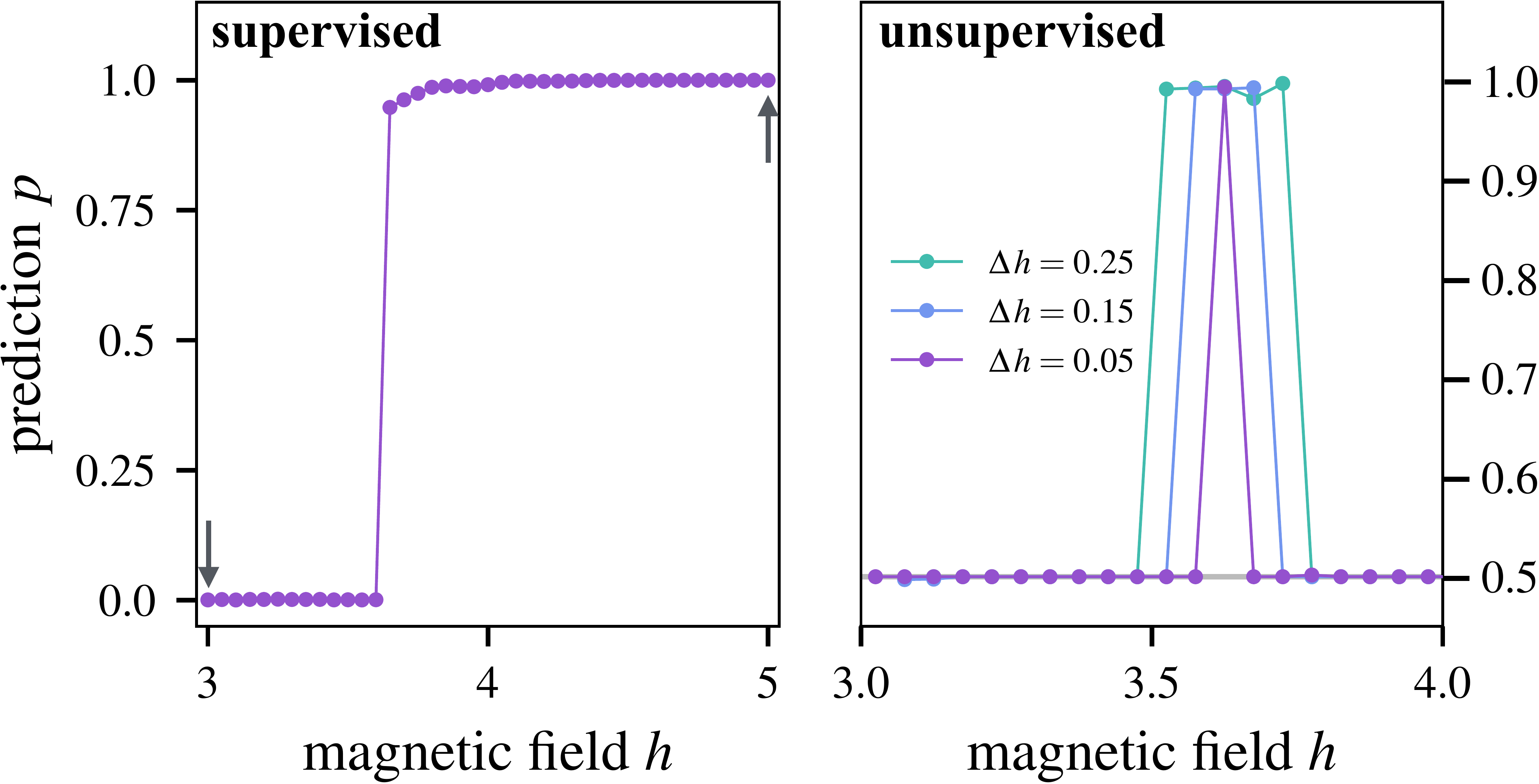}
	\caption{Identification of the {\bf first-order quantum phase transition} between the checkerboard solid and superfluid phase
			for $\Delta=3$ employing (a) a supervised and (b) the unsupervised ML approach.
	\label{fig:qpt_cut}}
\end{figure}

\begin{figure*}[t]
	\includegraphics[width=\textwidth]{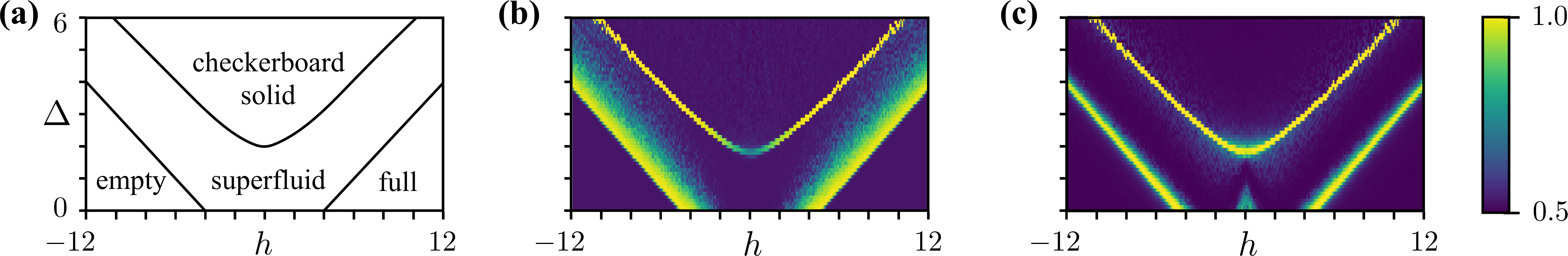}
	\caption{ {\bf Zero-temperature phase diagram of interacting hardcore bosons}.
			Panel (a) shows the phase diagram extracted from quantum Monte Carlo (QMC) simulations
			\cite{Batrouni2000,Schmid2002}.
			Panels (b) and (c) show the phase diagrams extracted from our unsupervised ML approach
			applied to correlation functions sampled in QMC simulations (for $L=8$).
			Panel (b) is based on the diagonal correlation function $\langle S_i^z S_j^z\rangle$ and
			panel (c) on the off-diagonal correlation function $\langle S_i^+ S_j^-\rangle + \langle S_j^+ S_i^-\rangle$, respectively.
	\label{fig:quantum_phase_diagram}}
\end{figure*}
\begin{figure*}[t]
	\includegraphics[width=0.95\textwidth]{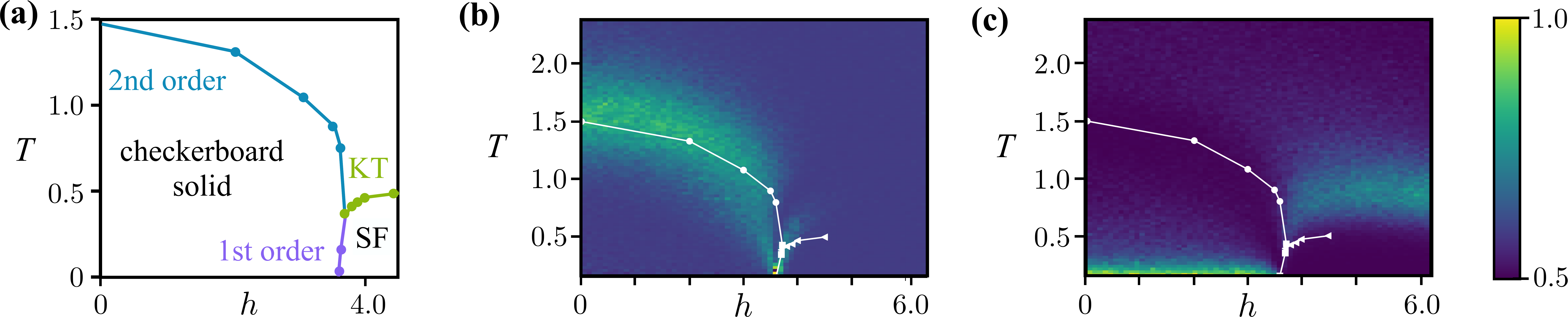}
	\caption{ {\bf Finite-temperature phase diagram of interacting hardcore bosons}.
			Panel (a) shows the phase diagram found in quantum Monte Carlo simulations \cite{Schmid2002},
			panels (b) and (c) show results form our unsupervised ML approach (for $L=8$).
			For Panel (b) the diagonal correlation function $\langle S_i^z S_j^z\rangle$ is fed into the CNN,
			for panel (c) the winding number per site.
			The white lines indicate the phase boundaries of panel (a).
	\label{fig:ft_phase_diagram}
	}
\end{figure*}
%


\noindent
{\em Hardcore bosons.--}
We put our unsupervised approach to practice by studying a prototypical quantum many-body system -- the Bose-Hubbard
model \cite{Fisher1989}, which captures the competition between kinetic and potential energies for bosons that stabilize
superfluid and Mott insulating phases, respectively. Adding nearest-neighbor repulsion or ring exchange terms can enrich its
phase diagram by supersolid \cite{Wessel2005,Heidarian2005,Melko2005} or $d$-wave-correlated Bose liquids \cite{Sheng2008}.
Here we restrict ourselves to a model of hardcore bosons on the square lattice \cite{Schmid2002} subject to a nearest-neighbor Coulomb repulsion $V$ and a chemical potential $\mu$
\begin{equation}
   \mathcal{H} = -t \sum_{\langle i,j \rangle} \left( a_i^\dagger a_j^{\phantom{\dagger}} +  a_j^\dagger a_i^{\phantom{\dagger}} \right)
   			+ V \sum_{\langle i,j \rangle} n_i n_j - \mu \sum_i n_i \,,
   \label{eq:BHM}
\end{equation}
where $n_i = a_i^\dagger a_i^{\phantom{\dagger}}$ are the usual boson operators in second quantization.
This relatively simple model exhibits a ground-state phase diagram \cite{Batrouni2000,Schmid2002} with four different phases
as illustrated in Fig.~\ref{fig:quantum_phase_diagram}a)
-- besides the trivial, fully filled or completely empty, ground states there is an extended superfluid phase  along with a checkerboard solid.
At finite temperatures, the model exhibits continuous phase transitions to a normal fluid both from the checkerboard solid (second order transition) and the superfluid (Kosterlitz-Thouless transition) as illustrated in Fig.~\ref{fig:ft_phase_diagram}a).
This plethora of phases and different types of phase transitions is ideally suited to benchmark our ML approach against numerically exact results
from large-scale Monte Carlo simulations \cite{Schmid2002}. In doing so, we map Hamiltonian \eqref{eq:BHM} to an anisotropic spin-1/2 model in a magnetic field \cite{Matsubara1956}
\begin{equation}
   \mathcal{H} = - \sum_{\langle i,j \rangle} \left( S_i^+ S_j^- +   S_i^- S_j^+ \right) + \Delta \sum_{\langle i,j \rangle} S_i^z S_j^z + h \sum_i S_i^z
   \label{eq:SpinModel}
\end{equation}
with $\Delta  = V$, $h=2V - \mu$ for $t =1$ and employ  a stochastic series expansion \cite{Sandvik1999}
to sample systems of linear system sizes $L=8,16,24,32$ down to inverse temperatures of $\beta=40$.


Starting with the ground-state phase diagram in the $(\Delta,h)$ plane we map out the phase boundaries using our unsupervised approach
by shifting a training window (of width $\delta_\Delta = 0.2$) vertically across the parameter space.
In the spirit of devising a rather general algorithm that uses no prior knowledge about the phase diagram and the specific nature of its phases,
we feed the CNN with equal-time Green's functions sampled in the Monte Carlo simulation, i.e. we rely on the CNN to extract sufficient information from this essential, but rather generic observable to discriminate different phases \cite{Broecker2016}.
In the context of Hamiltonian \eqref{eq:SpinModel}, we alternatively consider both the diagonal correlation function $\langle S_i^z S_j^z\rangle$ and off-diagonal correlation function $\langle S_i^+ S_j^-\rangle + \langle S_j^+ S_i^-\rangle$ as input.
Fig.~\ref{fig:qpt_cut} 
illustrates results for an example cut at $\Delta=3$ where we show the discrimination of superfluid versus checkerboard solid in both
a supervised and unsupervised learning approach.
In Fig.~\ref{fig:qpt_cut}a) we show that supervised learning deep in the two phases $(h_{1,2} = 3.0,5.0)$ allows to identify
the location of the phase transition via the change of the prediction function for intermediate values of $h$.
In Fig.~\ref{fig:qpt_cut}b) we show results from the unsupervised scheme put forward in this manuscript where we move training windows
of varying length across the cut. A singular peak in the average prediction success clearly indicates the location of the
phase transition, with the peak narrowing for shorter window width as expected.
Results from our ML approach for the entire phase diagram are given in Figs.~\ref{fig:quantum_phase_diagram}b) and c) where
we plot the average prediction success that reveals several sharp transitions and in fact traces out the phase diagram in superb
quantitative agreement with the original Monte Carlo analysis \cite{Schmid2002}. The minor broadening of the transition from one 
of the trivial states into the superfluid  in the diagonal correlation function reflects its slower decay in comparison with the rapid change
of the off-diagonal correlation function at the same transition (for a finite system size).

\begin{figure}[b]
	\includegraphics[width=0.9\columnwidth]{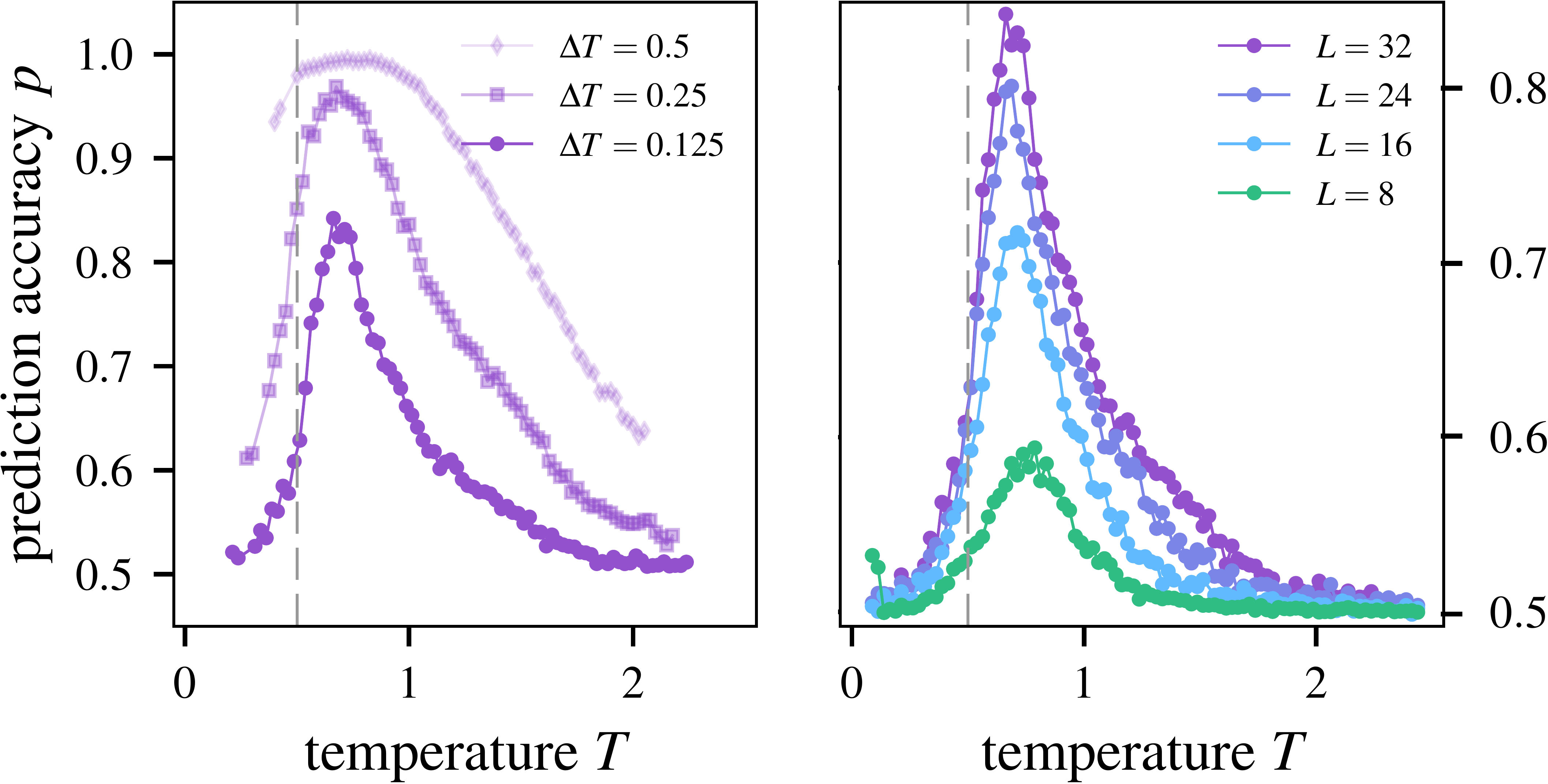}
	\caption{Identification of the  finite-temperature  {\bf Kosterlitz-Thouless transition} between superfluid and normal fluid
		     for $h=5.0$ when feeding the CNN with winding numbers of Monte Carlo configurations.
		     The peak in the prediction accuracy is located slightly above the actual location of the transition (dashed line).
		     }
	\label{fig:FiniteSize2}
\end{figure}
%


Turning to the finite-temperature phase diagram of model \eqref{eq:SpinModel} we find that tracing out the thermal phase
transitions with our ML approach is somewhat harder. 
Fig.~\ref{fig:ft_phase_diagram}b) shows the phase diagram extracted via our unsupervised approach when feeding the 
diagonal correlation function $\langle S_i^z S_j^z\rangle$ into the CNN. 
The second-order transition between checkerboard solid and normal fluid results in a relatively broad signature, 
which is mostly due to the moderate system size ($L=8$) underlying this comprehensive sweep of the phase diagram.
While the Kosterlitz-Thouless (KT) transition out of the superfluid leaves no visible trace in our ML analysis of the diagonal 
correlation function, see Fig.~\ref{fig:ft_phase_diagram}b), it leaves a broad signal in the off-diagonal correlation function (not shown).
This reflects the intrinsic inefficiency of local observables to capture the non-local nature of the vortex-antivortex unbinding at a KT transition.
Alternatively, we can feed the CNN with explicit information about winding numbers for configurations sampled in the Monte Carlo simulation, 
e.g. the winding number per site in one of the spatial directions~\footnote{
We normalize each sample of the winding number with respect to its largest absolute value to ensure values in the range of $\left(-1, 1\right)$.
The idea is to emphasize the {\em pattern} of the distribution of windings 
rather than the absolute values, which would be used for a direct estimation of the order parameter.}.
This results in a clear signal located slightly above the actual KT transition, see Fig.~\ref{fig:ft_phase_diagram}c).
The finite-size trend of this peak is shown in Fig.~\ref{fig:FiniteSize2}.
While the feature broadens with increasing system size, the peak systematically enhances for larger systems 
and slowly shifts down towards the Monte Carlo estimate.


\noindent
{\em Fermions and topological order.--}
Our second principal example is a model of Dirac fermions coupled to a fluctuating $Z_2$ gauge field
that exhibits a phase transition from a deconfined, topologically ordered phase to a conventional
antiferromagnetically ordered phase \cite{Assaad2016,Gazit2017}.
Its Hamiltonian is defined on a square lattice and reads
\begin{eqnarray}
   \mathcal{H} = & &
   \sum_{\langle i,j \rangle} Z_{\langle i,j \rangle} \left( \sum_{\alpha=1}^N c^\dagger_{i,\alpha} c^{\phantom{\dagger}}_{j,\alpha} + h.c \right)
   - N h \sum_{\langle i,j \rangle} X_{\langle i,j \rangle}  \nonumber \\
  & & + N F  \sum_{\square} \prod_{\langle i, j \rangle \in \partial \square}  Z_{\langle i,j \rangle} \,,
   \label{eq:Fermions}
\end{eqnarray}
where we consider $N=2$ species of fermions with creation/annihilation operators $c^\dagger_{i,\alpha}$/$c^{\phantom{\dagger}}_{i,\alpha}$
and bond spin operators $Z_{\langle i,j \rangle}$ and $X_{\langle i,j \rangle}$ that correspond to the usual Pauli spin-1/2 matrices.
Since
\begin{equation}
Q_i =  (-1)^{\sum_{\alpha} c^\dagger_{i,\alpha} c^{\phantom{\dagger}}_{i,\alpha} }  \prod_{\delta = \pm a_x, \pm a_y}  X_{\langle i, i + \delta \rangle}
\end{equation}
commutes with the  Hamiltonian, the Gauss law,  $Q_i =-1$, is imposed dynamically   in the zero temperature limit and on any finite sized lattice.
Here we have supplemented the original model of Ref.~\cite{Assaad2016} with a flux term of magnitude $F=1/2$. For this value of the flux,
the transition between the two aforementioned phases is driven by the strength of the magnetic field $h$ with the critical value
estimated to be $h_c \approx 0.40$ \cite{Assaad2017}. 

\begin{figure}[t]
	\includegraphics[width=0.98\columnwidth]{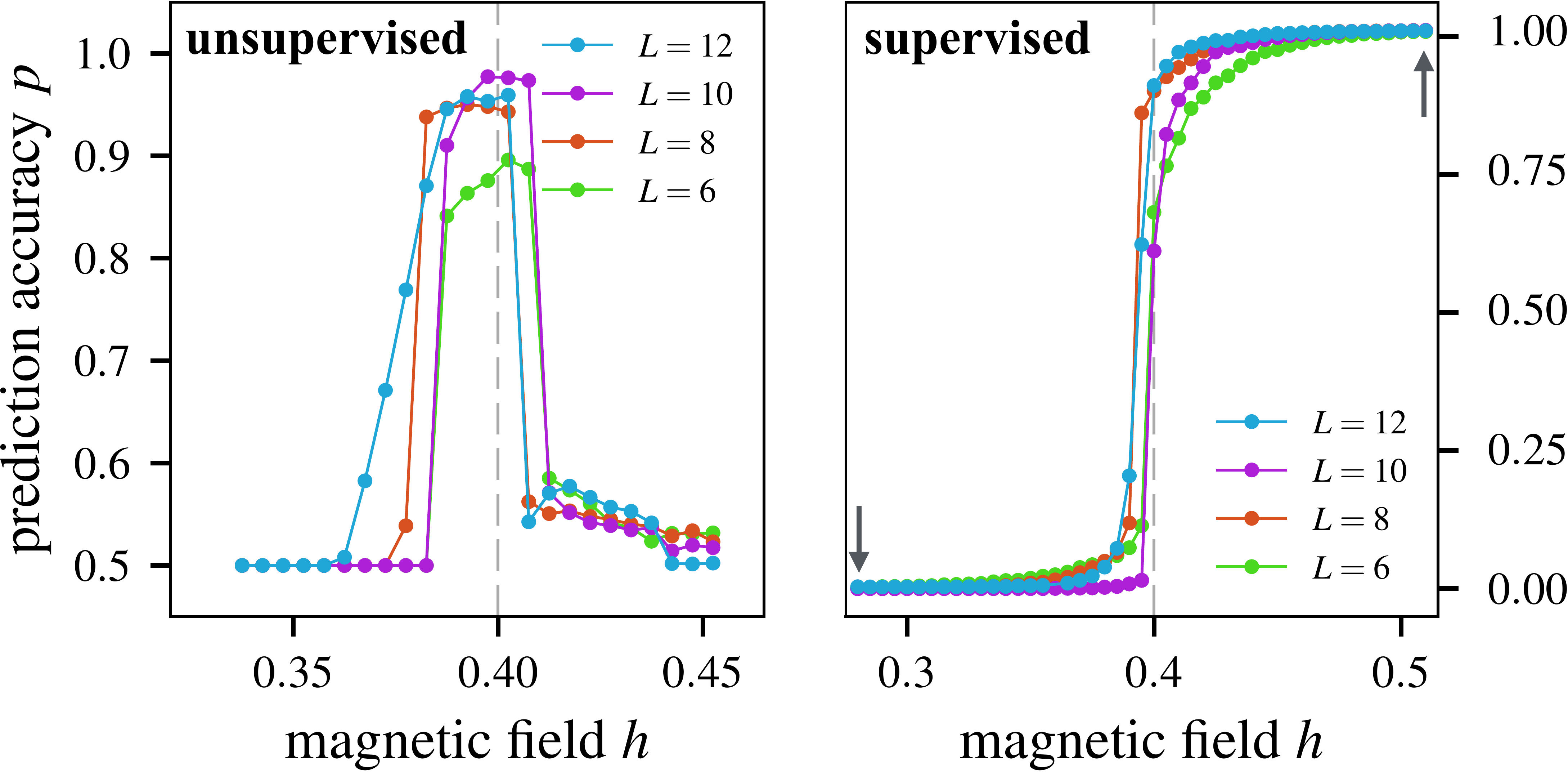}
	\caption{Detection of {\bf phase transition to topological order} in model \eqref{eq:Fermions}
		     of  fermions coupled to a fluctuating Z$_2$  gauge field 
		     employing (a) the unsupervised and (b) a supervised ML approach.}
	\label{fig:topo}
\end{figure}

We explore this model by combining our unsupervised ML approach with finite temperature auxiliary-field quantum Monte Carlo \cite{Gubernatis2016}  as implemented in the ALF-package  \cite{ALF_v1,Assaad08_rev} with the latter  providing samples of the equal-time single-particle Green's function $\langle c^\dagger_i c_j \rangle$ to the CNN \cite{Broecker2016} (for an inverse temperature $\beta  = 40$).
As Fig.~\ref{fig:topo}a) clearly demonstrates, the highly non-trivial phase transition in model \eqref{eq:Fermions} can be readily located using
our unsupervised approach -- there is a sharp peak located right at the expected value of the transition for varying system sizes.
This might be surprising at first sight as one might expect that the non-local nature of the topologically ordered phase might pose
similar problems as the identification of the vortex-antivortex unbinding at a topological phase transition (as discussed above).
Indeed, a recent ML-based identification of topological order \cite{Zhang2017} succeeded only because of the addition of
explicit non-local filters (akin to the convolutional filter of a CNN).
In the context of model \eqref{eq:Fermions} such steps are not necessary as the topological nature of the deconfined Dirac phase
can reveal itself already on relatively modest length scales --  the proliferation of vison excitations at the transition are  bound
to plaquettes of the square lattice and as such easily detectable. Although  visons  are very local, the CNN is fed  with  snapshots of the  Green's function $\langle c^\dagger_{i,\sigma} c_{j,\sigma} \rangle$ -- a quantity that, taken at face value, contains very little information. Since the simulations are SU(2)-spin invariant, each snapshot also has no spin dependence. Furthermore, since  $\left\{ Q_i, c_{i,\sigma} \right\} = 0 $, we have $\langle c^\dagger_{i,\sigma} c_{j,\sigma} \rangle = \delta_{i,j}/2 $, reflecting the fact that the Green's function is a {\em gauge-dependent} quantity. Note that the latter equation holds only after averaging over snapshots.  Given this background,  it   is certainly remarkable to see that the CNN can  detect   in such a precise manner the aforementioned phase  transition between  a topologically ordered state and an antiferromagnet.

As a consistency check we also show results from a supervised learning approach in Fig.~\ref{fig:topo}b) where we
have trained a CNN deep inside the two phases (indicated by the arrows) and observe that the prediction changes from 0 to 1 right
at the expected location of the transition. Note that both approaches  as well as standard analysis of the phase transition  using RG-invariant quantities \cite{Assaad2016,Assaad2017} are relatively sensitive to finite-size effects.
This certainly makes it hard to infer the order of the phase transition from the current data.
Compared to the hard first-order transition in the boson model, the fermionic transition at hand certainly does not show a similarly
sharp transition. On the other hand, the finite-size trends of Fig.~\ref{fig:topo} do not readily allow for a data collapse similar
to what has been demonstrated for the Ising model \cite{Carrasquilla2017}. 


\noindent
{\em Discussion.--}
In the recent surge of applying machine learning techniques to statistical physics problems, alternative unsupervised learning schemes \cite{Wang2016,Hu2017,Nieuwenburg2017,Liu2017b} have been tested on (classical) many-body problems. One prominent unsupervised approach
is the principal component analysis (PCA), which has been demonstrated \cite{Wang2016,Hu2017} to locate the phase transition of classical spin
models via a clustering analysis that correctly discriminates the formation of spatial ordering patterns and symmetry breaking from disordered phases. In such relatively simple scenarios, the dominant principle component in fact reflects the order parameter of the phase. However, it remains to be seen whether the PCA is similarly suitable to quantum many-body systems that allow for considerably more subtle forms of order -- such as the formation of superfluids or topological order as demonstrated for our approach.
Probably, the approach closest in spirit to ours is the ``learning by confusion" scheme put forward in Ref.~\onlinecite{Nieuwenburg2017}
(and expanded in a recent preprint \cite{Liu2017b}).
This scheme also discusses a way to turn an initially supervised learning approach into an unsupervised one by attempting various
splits of labeled instances in the training step. While this procedure allows to locate phase transitions via the identification of a local 
maximum in the learning success, we find that our current approach produces a considerably sharper signal of the phase transition.
Further refining our approach might even allow to independently extract the order of a phase transition from the form and finite-size
behavior of the peak, which we will address in future studies.


\acknowledgments
P.B. acknowledges support from the Deutsche Telekom Stiftung and the Bonn-Cologne Graduate School of Physics and Astronomy (BCGS).
The Cologne group was partially supported by the DFG within the CRC network TR 183 (project B01) and gratefully acknowledges support
from the Quantum Matter and Materials (QM$^{\text{2}}$) initiative at Cologne.  
F.F.A. is partially supported by the German Research Foundation (DFG)  FOR-1807 grant. 
The numerical simulations were performed on the CHEOPS cluster at RRZK Cologne
and the JURECA cluster at the Forschungszentrum J\"ulich.


\bibliography{ML}

\end{document}